\documentclass[prd,twocolumn,preprintnumbers,superscriptaddress,nofootinbib]{revtex4-2}
\usepackage[a4paper, hdivide={1.91cm,,1.165cm}, vdivide={1.83cm,,3.6cm}]{geometry}

\usepackage{amstext,amssymb}
\usepackage{amsmath}
\usepackage{graphicx}
\usepackage[hyperfootnotes=true]{hyperref}
\usepackage{xspace}
\usepackage{color}
\usepackage{units}
\usepackage{slashed} 

\usepackage{longtable}
\usepackage{multirow}

\usepackage{wrapfig}
\definecolor{light-gray}{gray}{0.95}
\usepackage{tcolorbox}
\newcommand{\bmt}{\begin{pmatrix}}
\newcommand{\emt}{\end{pmatrix}}

\begin{document}

\title{CDF II W-mass anomaly and SO(10) GUT}
%
\author{Purushottam \surname{Sahu}}
\email{purushottams@iitbhilai.ac.in}
\affiliation{Department of Physics, Indian Institute of Technology (IIT) Bhilai, Raipur 492015, India}
\affiliation{International Centre for Theoretical Physics (ICTP), Strada Costiera 11, Trieste 34151, Italy}
\affiliation{Department of Physics, Indian Institute of Technology Bombay,
Powai, Mumbai, Maharashtra 400076, India}

\author{Hiranmaya \surname{Mishra}}
\email{hm@prl.res.in}
\affiliation{School of Physical Sciences, National Institute of Science Education and Research (NISER) Bhubanwswar, HBNI, Jatni 752050,Odisha, India}
\author{Prasanta K. \surname{Panigrahi}}
\email{panigrahi.iiser@gmail.com}
\affiliation{Indian Institute of Science Education and Research (IISER) Kolkata, India}
\author{Sudhanwa \surname{Patra}}
\email{sudhanwa@iitbhilai.ac.in}
\affiliation{Department of Physics, Indian Institute of Technology (IIT) Bhilai, Raipur 492015, India}
\author{Utpal \surname{Sarkar}}
\email{utpal.sarkar.prl@gmail.com}
\affiliation{Indian Institute of Science Education and Research (IISER) Kolkata, India}

\begin{abstract}
The W-mass anomaly has yet to be established, but a huge proliferation of articles on the subject established the rich potential of such event. We investigate the SO(10) GUT constraints from the recently reported W-mass anomaly. We consider both Supersymmetric (SUSY) and non-supersymmetric (non-SUSY) grand unified theories by studying renormalization group equations (RGEs) for gauge coupling unification and their predictions on proton decay.  
In the non-SUSY models, single-stage unification is possible if one include a light (around TeV) real triplet Higgs scalar. However, these models predict speedy proton decay, inconsistent with the present experimental bound on the proton decay. This situation may be improved by including newer scalars and new intermediate-mass scales, which are present in the $SO(10)$ GUTs. The standard model is extended to a left-right symmetric model (LR), and the scale of LR breaking naturally introduces the intermediate scale in the model. A single-stage unification is possible even without including any triplet Higgs scalar in a minimal supersymmetric standard model. 
\end{abstract}
\pacs{98.80.Cq,14.60.Pq} 
\maketitle 
\section{Introduction} 
\label{sec:intro}
The Standard Model (SM) of Particle Physics, one of the particle physics community's greatest treasures, has successfully explained nearly all experimental data up to the current accelerator energy scale. At the same time, it fails to answer theoretical concerns such as the origin of non-zero neutrino masses, the matter-antimatter asymmetry of the universe, dark matter and dark energy, and so on, suggesting that SM is not the final theory. Alternatively, precision measurements of observables may be critical in testing the SM and shedding information on the possibility of Beyond Standard Model (BSM) physics. 
The CDF II collaboration's most recent precision measurement indicated a shift in W boson mass~\cite{CDF:2022hxs}: $$m^{\rm CDF}_W = \big(80.433 \pm 0.0064_{\rm stat} \pm 0.0069_{\rm syst} \big)\mbox{GeV}\,.$$ 
This CDF II data clearly shows disagreement with SM prediction and is also in align with the earlier global data from LEP, CDF, D0, and ATLAS, with mass range of W boson mass, $m_W = (80.357\pm 0.006)$~GeV~\cite{ParticleDataGroup:2018ovx}.

Many proposals have been investigated to account for the shift in W boson mass; the most straightforward extension among them is to incorporate a real scalar triplet without any hypercharge that explains the requisite shift in W boson mass without impacting the Z boson mass.
Although Ref~\cite{Strumia:2022qkt} made the initial proposal to explain CDF W mass anomaly by extending SM with a real scalar triplet with zero hypercharge, the detailed phenomenology of scalar triplet with zero hypercharge in the context of CDF II W boson mass anomaly was studied in Ref~\cite{FileviezPerez:2022lxp}. Interestingly, the relation of the W-boson mass anomaly to grand unified theories such as SU(5) has been investigated by adding complex scalar triplets without hypercharge~\cite{Evans:2022dgq,Senjanovic:2022zwy,Shimizu:2023rvi}. In~\cite{Evans:2022dgq} they claimed that $SU(5)$ GUT with minimal representations of $24_H, 5_H, \overline{5}^a_{F}, 10^a_F, \overline{10}^a_F$ can address the W boson mass anomaly while replicating SM fermion masses and mixings. Another group performed a similar analysis within SU(5) GUT by adding scalar triplet and fermion triplet without hypercharge~\cite{Senjanovic:2022zwy}, where they consider representations such as $24_H, 24_F, 5_H, \overline{5}^a_{F}, 10^a_F$. Extending on the ideas of explaining the W-boson mass anomaly and its connection to grand unified theories, we intend to investigate $SO(10)$ GUT with and without any intermediate symmetry-breaking steps, with implications for gauge coupling unification, experimental constraints on proton decay, neutrino mass, and the universe's matter-antimatter symmetry, among other things. We explore a non-supersymmetric $SO(10)$ GUT with direct breaking to SM and explain the W-boson mass anomaly by introducing an additional scalar degree of freedom at the TeV scale. For the analysis, the SO(10) representations are $SO(10): 10_H; \quad 16^a_F; \quad 45_H$ for direct breaking of $SO(10)$ to SM and $SO(10): 10_H; \quad 16^a_F; \quad 126_H; \quad 45_H$ for an intermediate Left-right symmetric models (LRSM) breaking between SO(10) and SM.


Left-right symmetric models ~\cite{Mohapatra:1974gc,Senjanovic:1975rk,Senjanovic:1978ev,Mohapatra:1980yp} are believed to be one of the promising extensions of the SM that offer a reasonable explanation for parity violation. In such models, the gauge symmetry is improved to $SU(3)_c \times SU(2)_L \times SU(2)_R \times U(1)_{B-L}$, with left-handed leptons and quarks translating as $SU(2)_L$ doublets with right-handed counterparts as $SU(2)_R$ doublets. Due to the inclusion of right-handed neutrinos to complete the gauge multiplets, these models, not only shed light on the cause of parity violation but also explain small neutrino masses using the type-I \cite{Minkowski:1977sc, Mohapatra:1979ia, Yanagida:1979as, GellMann:1980vs} or type-II 
\cite{Mohapatra:1980yp,Schechter:1980gr,Cheng:1980qt, Magg:1980ut} seesaw mechanism. Furthermore, they provide a mathematical description of fermion hypercharge quantum numbers derived from baryon number minus lepton number (B-L) charges and the third component of the right-handed isospin. In this letter, we investigate at the scenario of the $SO(10)$ GUT, which not only accommodates CDF II W boson anomaly, but also able to unify all matter fields (including the right-handed neutrino) of each generation in a single $SO(10)$ representation, for example, the spinorial $16_F$.

The paper is organized as follows: In section II, We briefly introduce the explanation of the CDF W mass anomaly by extending SM using the isospin $SU(2)_L$ triplet scalar to make this study self-contained. In section III, we examine a novel symmetry breaking of non-supersymmetric SO(10) GUT without having any intermediate symmetry, along with including a $SU(2)_L$ scalar triplet with zero hypercharge from the electroweak scale onwards to accommodate the W mass anomaly, as well as to study the evolution of the gauge coupling constants of SM gauge symmetry and proton decay predictions. In section IV: to address proton decay constraints and accommodate W mass anomaly, we embed the framework in a non-SUSY SO(10) grand unified theory with an intermediate left-right symmetry potential to explain neutrino mass and the matter-antimatter asymmetry of the universe. 
We discuss the inclusion of supersymmetric SO(10) GUT with and without intermediate left-right symmetry in section V.  
Finally, we summarize our significant findings and scope for future studies.

\section{CDF W mass anomaly with SM plus isospin $SU(2)_L$ triplet scalar} 
It has been pointed out that the recent CDF W boson mass shift from SM predictions might arise due to the presence of complex zero hypercharge triplet Higgs boson $\Omega_{1_C, 3_L, 0_Y}$~\cite{Athron:2022isz, FileviezPerez:2022lxp}. The tiny mass shift of the W boson mass can be interpreted as sub-dominant contribution arising from the small induced VEV of the same scalar triplet. 
The relevant Lagrangian involving SM Higgs $H$ and triplet scalar $\Omega$ is given by
\begin{eqnarray}
\mathcal{L} = \big(D_\mu H \big)^\dagger \big(D^\mu H \big) 
+ {\rm Tr}\left[ (D_\mu \Omega)^\dag (D^\mu \Omega) \right] - V(H, \Omega), \nonumber 
\\
\label{eq:triplet}
\end{eqnarray}
where the SM Higgs field is denoted by $H^T= \big( \phi^+, \phi^0 \big)$ and the real scalar triplet $\Omega$ in its matrix representation as,
\begin{equation}
 \Omega = \frac{1}{2} \begin{pmatrix}
                       \Omega^0  & \sqrt{2} \Omega^+ \\
                       \sqrt{2} \Omega^- & -\Omega^0
                      \end{pmatrix} .
\end{equation}
The covariant derivative involving SM Higgs $H$ is straightforward, while for the scalar triplet, it is $D_\mu \Omega = \partial_\mu \Omega - i g_{2L} [W_\mu, \Omega]$ where $\Omega = \frac{\tau^a}{2} \Omega^a$ and $W_\mu = \frac{\tau^a}{2} W_\mu^a$ with $\tau^a, a=1,2,3$ the Pauli matrices. The $W_\mu$ ($g_{2L}$) is the corresponding gauge boson (gauge coupling) of the $SU(2)_L$ gauge group. After spontaneous symmetry breaking, we get known SM Higgs bosons, one CP-even heavy Higgs boson, a CP-odd Higgs boson, and two charged Higgs bosons. The relevant scalar potential involving SM Higgs boson $H$ and scalar triplet $\Omega$ is presented below 
\begin{eqnarray}
 V(H, \Omega) &=& - \mu_H^2 H^\dagger H  + \lambda_H \big( H^\dagger H\big)^2 
 - M^2_\Omega {\rm Tr}( \Omega^2) \nonumber \\
 &&+ \lambda_{\Omega} {\rm Tr}( \Omega^4)
 + \lambda^\prime_{\Omega} \big({\rm Tr} \Omega^2\big)^2
+ \mu_{H \Omega} H^\dag \Omega H \nonumber \\
  &&+ \alpha \big( H^\dagger H\big) {\rm Tr}( \Omega^2) 
  + \beta H^\dag \Omega^2 H
  + h.c.\label{eq:Higgs_poten}
\end{eqnarray}
where $\mu_{H \Omega}$ is considered to be real. The spontaneous symmetry breaking is achieved by assigning non-zero VEV to SM Higgs as well as to the scalar triplet. Thus, after spontaneous symmetry breaking, the modified Higgs fields are defined as
\begin{eqnarray}
&&H = \begin{pmatrix}
       \phi^+ \\
       \frac{v+h+i A}{\sqrt{2}}
     \end{pmatrix}
\nonumber \\
&& \Omega = \frac{1}{2} \begin{pmatrix}
                       v_\Omega + \Omega  & \sqrt{2} \Omega^+ \\
                       \sqrt{2} \Omega^- & - v_\Omega -\Omega
                      \end{pmatrix}
\end{eqnarray}

After simplifications, the resulting mass formula for CP-even, CP-odd, and charged Higgs is, 
\begin{eqnarray}
&&M^2_{h} =  2 \lambda_H v^2 \nonumber \\
&&M^2_\omega = 2 \rho v^2_\Omega + \frac{\mu_{H \Omega} v^2}{4 v_\Omega}
 \nonumber \\
&&M^2_{\omega^{\pm}} = \mu_{H \Omega} v_\Omega \bigg(1+\frac{v^2}{4 v^2_\Omega} \bigg)
\end{eqnarray}
where $\rho = \lambda_\Omega + \lambda^\prime_\Omega$. The physical fields $(h,\omega)$ are related to $(\phi^0, \Omega^0)$ by some rotation mixing matrix.
The shift in W boson mass can be explained using the extra contribution arising from scalar triplet VEV
$$M_W = M^{\rm SM}_W + g^2_{2L} v^2_\Omega $$ ,
due to the presence of interaction $\Omega^0 W^-_\mu W^{+ \mu}$. Taking $g_{2L} (\mu = M_Z) = 0.65171$, we estimate the value of $v_\Omega$ as $5.4$~GeV.

\section{CDF W boson mass anomaly and non-SUSY $SO(10)$ GUT}
\label{sec:SO10:Wmass}
In the previous section, it is evident that the shift in mass of W boson can be explained if SM is extended with a scalar triplet with zero hypercharge, lying around TeV scale~\footnote{The experimentally reported CDF II W boson mass anomaly has been nicely accommodated in $SU(5)$ GUT in ref~\cite{Evans:2022dgq} with the inclusion of scalar triplet $\Omega (1_C, 3_L, 0_Y)$ at the electroweak scale. With representations like $\{24_H,~5_H,~\overline{5}^a_{F},~10^a_F,~ \overline{10}^a_F\}$ one can explain charged fermion masses and mixing along with accommodating W mass anomaly. However, the model predicts massless neutrinos, which contradicts neutrino oscillation data. Other proposal~\cite{Senjanovic:2022zwy} has been put forward where SM is extended with both scalar triplet $\Omega (1_C, 3_L, 0_Y)$ and fermion triplet $\Sigma (1_C, 3_L, 0_Y)$ having hypercharge zero. However, the proposal requires large number of $SU(5)$ representations $\{24_H,~ 24_F,~5_H,~\overline{5}^a_{F},~10^a_F\}$ to explain non-zero neutrino mass.}. We consider a novel symmetry breaking of non-supersymmetric $SO(10)$ GUT~\cite{Pati:1974yy,Fritzsch:1974nn,PhysRevD.48.264,Senjanovic:1975rk,Clark:1982ai,Chang:1984qr,Bertolini:2009qj,Altarelli:2013aqa,Dueck:2013gca,Meloni:2014rga,Meloni:2016rnt,Preda:2022izo}, without having any intermediate symmetry and adding a $SU(2)_L$ scalar triplet with zero hypercharge at electroweak scale onwards for accommodating W boson mass anomaly as,
\begin{eqnarray}
SO(10)
&&	\stackrel{M_U}{\longrightarrow} SU(3)_C\otimes SU(2)_L\otimes U(1)_{Y} \nonumber \\
&&	\stackrel{M_Z}{\longrightarrow} SU(3)_C\otimes U(1)_{Q}\;.
\label{Chain}
\end{eqnarray}
The $SO(10)$ breaks down to SM by assigning a non-zero VEV to the singlet direction of $45_H$, while the subsequent symmetry breaking of SM to low energy theory is achieved by SM Higgs contained in $10_H$. All the SM fermions per generation plus a right-handed neutrino are unified within spinorial representation $16_F$ of $SO(10)$. Interestingly, the minimal addition of $\Omega$ can also be found in $45_H$. Thus, we need one complex $10_H$, $45_H$ and $16^a_F$ for explanation of CDF W mass anomaly, light neutrino masses via type-I seesaw mechanism with the presence of right-handed neutrinos~\cite{Minkowski:1977sc, Mohapatra:1979ia, Yanagida:1979as, GellMann:1980vs, Babu:1992ia}: 

\[ 10_H \equiv 
H_{\Phi}\left(10\right)=\Phi(1,2,2;0)\oplus(3,1,1;-\frac{1}{3})\oplus(\overline{3},1,1;\frac{1}{3})\,.\]

All the SM gauge bosons ($G^{a}_\mu$, $W^i_{\mu}$ and $B_\mu$) plus additional vector bosons are contained in a 45-dimensional representation of SO(10), out of which a few vector boson can mediate proton decay. 
The SM fermions plus a right-handed neutrino are contained in a spinorial representation of $SO(10)$ as,
\begin{eqnarray*}
16_F &=& Q_L(3,2,1/6) \oplus u_R (3,1,2/3) \oplus d_R(3,1,-1/3) \\
 &&  \oplus \ell_L (1,2,-1/2) \oplus e_R (1,1,-1) \oplus N_R (1,1,0)\,. \quad (\mbox{SM})
  \end{eqnarray*}
 

The relevant $SO(10)$ invariant Higgs potential, involving $10_H \equiv \Phi_H = H$ and $45_H \equiv A(45)$, is written as
\begin{eqnarray*}
V_{SO(10)}&&\supset  \mu_{A}^{2}\, A_{ab}A_{ba}+\mu_{H}^{2}\, H_{a}H_{a}+ \lambda_{A}\, A^{2}A^{2}+\lambda'_{A}A^{4} \\
 &&+\lambda_{H}\, H^{4} + H_{a} g_{HA}A_{ab}A_{bc} H_{c}+g'_{HA}A^{2} H^{2} .\end{eqnarray*}

\begin{figure}[htb!]
\centering
\includegraphics[scale=0.50]{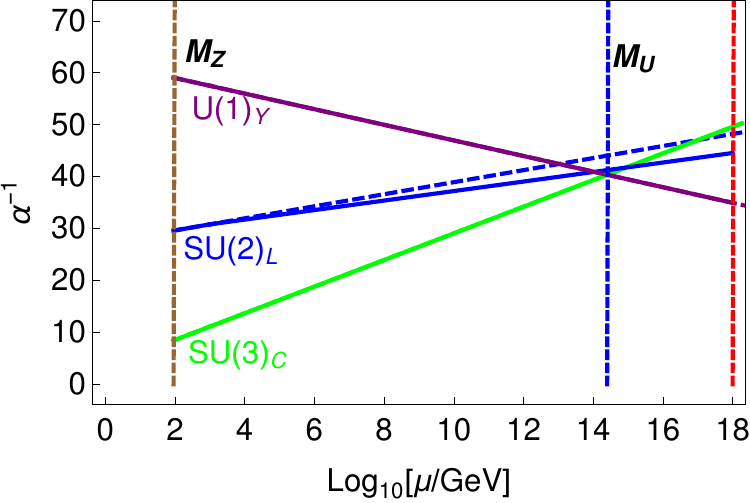}
\hspace{0.2 cm}
\caption{The behavior of evolution of the gauge coupling constants of SM gauge symmetry by adding a scalar triplet with zero hypercharge (as in ref.\cite{Evans:2022dgq}).  The dashed lines are contributions from SM particle content, and solid lines correspond to RGEs with extra particles.
.}
\label{plot:unifi_wo_threshold}
\end{figure}

\begin{table}[h]
\begin{tabular}{l c c}
\hline\hline
Mass Range &   1-loop level  \\[1mm]
\hline
$M_Z-M_I$   &  $b_i = (-7, -\frac{19}{6}, \frac{41}{10})$ \\
$M_I-M_U$   &  $b_i^\prime = (-7, -\frac{19}{6}+\frac{2}{3}*n_\Omega, \frac{41}{10})$  ~\cite{Evans:2022dgq}\\[2mm]  
$M_I-M_U$   &  $b_i^\prime = (-7, -\frac{19}{6}+\frac{1}{3}*n_\Omega+\frac{2}{3}*n_\Sigma, \frac{41}{10})$ ~\cite{Senjanovic:2022zwy}\\  
\hline
\end{tabular}
\caption{Explanation of CDF W mass anomaly requiring an extension of SM with a TeV scale scalar (color singlet, weak isospin triplet, and zero hypercharge) and their beta coefficients at one-loop and two-loop levels for the study of RGEs of gauge coupling constants. The $n_\Omega$ is 2 for complex and 1 for real scalar triplet.}
\label{tab:beta}
\end{table}
The evolution of gauge couplings for $SU(3)_C$, $SU(2)_L$ and $U(1)_Y$ gauge groups are displayed in Fig.\ref{plot:unifi_wo_threshold}. Here, the dashed (solid) lines correspond to SM contribution (SM plus scalar triplet contributions). The purple line represents the inverse fine structure constant for $U(1)_Y$ group whereas blue and green lines indicate the $SU(2)_L$ and $SU(3)_C$ groups respectively. The SM predictions with dashed lines clearly show that there is no such gauge coupling unification. However, as more particles are added to the SM at the TeV scale, the evolution of gauge couplings begins to break from the SM results and give successful gauge coupling unification of weak, electromagnetic, and strong forces.

The gauge coupling constants of SM meet close to $M_U \simeq 10^{14.4}$~GeV at a two-loop level while one-loop results are close to $M_U$ as shown in Fig.(\ref{plot:unifi_wo_threshold}) and Ref.\cite{Evans:2022dgq}. On the other hand, the one-loop results are exact for RGE analysis, including both scalar and fermion triplet at few TeV scale~\cite{Senjanovic:2022zwy}.

%
The super heavy gauge bosons and scalars carrying fractional charges as well as nontrivial color quantum numbers can mediate various proton decay modes~\cite{Buras:1977yy,Ibanez:1984ni,Babu:1992ia,BhupalDev:2010he,Chakrabortty:2019fov,Bhattiprolu:2022xhm}. The experimental bound on proton lifetimes in various decay modes are displayed in Table\,\ref{tab:pdecay1}. We wish to calculate the proton lifetime in the context of $SO(10)$ GUT, breaking to SM directly accommodating CDF W boson mass and compare how the model predictions are closer to present bounds set by recent or planned experiments~\cite{Super-Kamiokande:2020wjk}. It has been established that the present best limits were set on proton lifetime by Super-Kamiokande (SK) experiments as presented in Table\,\ref{tab:pdecay1} in the decay modes $p \to e^+ \pi^0$ and $p \to \mu^+ \pi^0$. We have omitted here the discussion on the less important decay modes of proton decay mediated by super heavy color triplet scalars.  
Thus, we focus on the contributions arising from the  dimension-6 operators mediating the proton decay with the exchange of lepto-quark gauge bosons violating both baryon and lepton numbers simultaneously. 
The spontaneous symmetry breaking of $SO(10)$ to SM happens at grand unified scale $M_{\rm GUT} \equiv M_U$, also leading to masses for these lepto-quark gauge bosons at this scale. 

\begin{table}[h]
\centering
\vspace{-2pt}
\begin{tabular}{||c|c|||}
\hline \hline
 Decay Modes & Expt. Bound (yrs)  \\
\hline
\hline
 $p \to \pi^0 e^+$ & $> 2.4 \times 10^{34}$~\cite{Super-Kamiokande:2020wjk} \\
  $p \to \pi^0 \mu^+$ & $> 1.6 \times 10^{34}$~\cite{Super-Kamiokande:2020wjk} \\
  \hline
  $p \to K^0 e^+$ & $> 1.0 \times 10^{33}$~\cite{Super-Kamiokande:2022egr} \\
  $p \to K^0 \mu^+$ & $> 3.6 \times 10^{33}$~\cite{Super-Kamiokande:2013rwg}\\
  \hline
  $p \to \pi^+ \overline{\nu}$ & $> 3.9 \times 10^{32}$~\cite{Super-Kamiokande:2014otb} \\
  $p \to K^+ \overline{\nu}$ & $> 5.9 \times 10^{33}$~\cite{Super-Kamiokande:2014otb} \\
\hline \hline
\end{tabular}
\caption{Experimental limit on life time of the proton decay for its various decay channels using Super-Kamiokande experiment~~\cite{Super-Kamiokande:2020wjk,Super-Kamiokande:2014otb,Super-Kamiokande:2022egr,Super-Kamiokande:2013rwg}.}
\label{tab:pdecay1}
\end{table}
Following~\cite{Sahu:2022rwq} and using the input model parameters, beta coefficients presented in Table (\ref{tab:beta}) and solving the set of standard RGEs~\cite{Georgi:1974yf} for gauge coupling constants, the estimated values of unification mass scale and inverse GUT coupling constant are as follows,
$$M_U= 10^{14.40}\,\,\mbox{GeV and} \quad \alpha^{-1}_G=39.0\,.$$

The predicted values of $\tau_p$ for the above-mentioned unification mass scale and inverse GUT coupling constant, including scalar and fermion triplet at TeV scale within SO(10) GUT is presented in Table (\ref{thrshldcr}). This prediction is significantly below the current bound provided by Super-Kamiokande and Hyper-Kamiokande studies, as seen in Table (\ref{tab:pdecay1}).
\begin{table}[h!]
	\begin{center}
		\begin{tabular}{|c|c|c|c|}
			\hline
			\multirow{1}{*}{Breaking chain} & \multirow{1}{*}{Observables} & \multicolumn{2}{c|}{Model Predictions} \\
			\hline
			\multirow{3}{*}{$SO(10) \to SM$} 
			& $\log_{10}\big(\frac{M_U}{\rm{GeV}} \big)$ & $14.4$ & $14.6$  \\
			\cline{2-4}
			& $\alpha_U^{-1}$ & $39.0$ & $40.0$ 
			\\
			\cline{2-4}
			& $\tau_p$ (yrs) & $3.5\times 10^{30}$ & $1.1\times 10^{32}$ \\
			\hline
		\end{tabular}
		\caption{Prediction of proton decay lifetime $\tau_{p}$ within $SO(10)$ GUT.}\label{thrshldcr}
	\end{center}
\end{table}
We will see in the following section that allowing intermediate left-right symmetry between $SO(10)$ and SM will enable us to overcome the problem of proton decay limitations.
\section{Impact of W-mass anomaly in SO(10) GUT with intermediate left-right symmetry}
\label{sec:SO10-lrsm:Wmass}
We now examine the impact of a scalar triplet $\Omega (1_C,3_L,0_Y)$, included at electroweak scale for the explanation of the W boson mass anomaly, by embedding the framework in non-SUSY $SO(10)$ grand unified theory with intermediate left-right symmetry. We consider breaking of $SO(10)$ GUT symmetry to SM via intermediate left-right symmetry \cite{Mohapatra:1974gc,Senjanovic:1975rk,Senjanovic:1978ev,Mohapatra:1980yp} as Pati-Salam symmetry~\cite{Pati:1974yy,Pati:1973uk} $SU(4)_C \times SU(2)_L \times SU(2)_R$  or $SU(3)_C \times SU(2)_L \times SU(2)_R \times U(1)_{B-L}$. The important advantage of accommodating CDF W mass anomaly in $SO(10)$ GUT in comparison to $SU(5)$ GUT is that $SO(10)$ is a rank five group while SM is a rank four gauge group and thus, can allow intermediate symmetries between $SO(10)$ to SM which thereby can solve the proton decay constraints along with the potential to address non-zero neutrino masses \cite{Babu:1992ia}, matter-antimatter asymmetry of the universe \cite{Buchmuller:1996pa,Buccella:2001tq,Branco:2002kt,Nezri:2000pb,Akhmedov:2003dg,DiBari:2021fhs} and neutrinoless double beta decay etc.  

The $SO(10)$ GUT can be spontaneously broken down to left-right symmetry $\mathbb{G}_{3221} \equiv SU(3)_C \otimes SU(2)_L\otimes SU(2)_R\otimes U(1)_{B-L}$ at unification mass scale $M_U$. One can use Higgs multiplets belonging to the $45$ and $210$ representations for spontaneous breaking of $SO(10)$ GUT. The subsequent stage of symmetry breaking from $\mathbb{G}_{3221} \to \mathbb{G}_{SM}$ is done at an intermediate-mass scale $M_I$. The Higgs scalar $\Delta_R$ ($H_R$), contained in $126$ ($16$) representation of SO(10) GUT, can be used for breaking of LRSM to SM depending upon the exact phenomenology one wish to explore. 
The interesting point to note here is that the inclusion of intermediate symmetry breaking scale demands the existence of right-handed neutrinos and/or scalar triplets at $M_I$ and can explain light neutrino masses via well-known type-I (and/or, type-II) seesaw mechanism and matter-antimatter asymmetry of the universe via Leptogenesis. One can introduce the discrete left-right symmetry invariant in addition to the gauge symmetry to simplify model parameters required for fermion masses and mixing. The last stage of symmetry breaking is done with known SM Higgs contained in $10_H$. At this stage, we include a scalar triplet with zero hypercharge for CDF II W mass anomaly, and then, this field will be present from $M_Z$ scale onwards. This is an equally important, on a stage of symmetry breaking, simultaneously addressing SM fermion masses and mixing along with the mass shift in W boson mass.

The SM Higgs $\Phi$ contained in $10_{H}$ is required to break SM to low energy theory, which can reproduce fermion masses and mixing. The decomposition of 10-dimensional Higgs field $H_\Phi$ under Pati-Salam symmetry and left-right symmetry having $B-L$ gauge symmetry as,
\[ 10_H \equiv 
H_{\Phi}\left(10\right)=\Phi(1,2,2;0)\oplus(3,1,1;-\frac{1}{3})\oplus(\overline{3},1,1;\frac{1}{3})\,.\]

Under left-right gauge symmetry, all the SM fermions plus a right-handed neutrino are contained in a spinorial representation of $SO(10)$ as,
\begin{eqnarray*}
16_F & = & \ell_{L}(1,2,1,-1)\oplus \ell_{R}(1,1,2,1) \\
 &  & \oplus Q_L(3,2,1,\frac{1}{3})\oplus Q_R(\overline{3},1,2,-\frac{1}{3})\,. \quad (\mbox{LRSM})
  \end{eqnarray*}
Similarly, 
\begin{eqnarray*}
45_H \equiv A(45) & = & (1,1,1;0)\oplus \Omega_L(1,3,1;0)\oplus \Omega_{R}(1,1,3;0)\\
 &  & \oplus(3,1,1;\frac{4}{3})\oplus(\overline{3},1,1;-\frac{4}{3})\oplus(8,1,1;0)\\
 &  & \oplus(3,2,2;\frac{2}{3})\oplus(\overline{3},2,2;-\frac{2}{3})\,.\end{eqnarray*}

\begin{widetext}
\begin{figure*}[t!]
\centering
\includegraphics[scale=0.45]{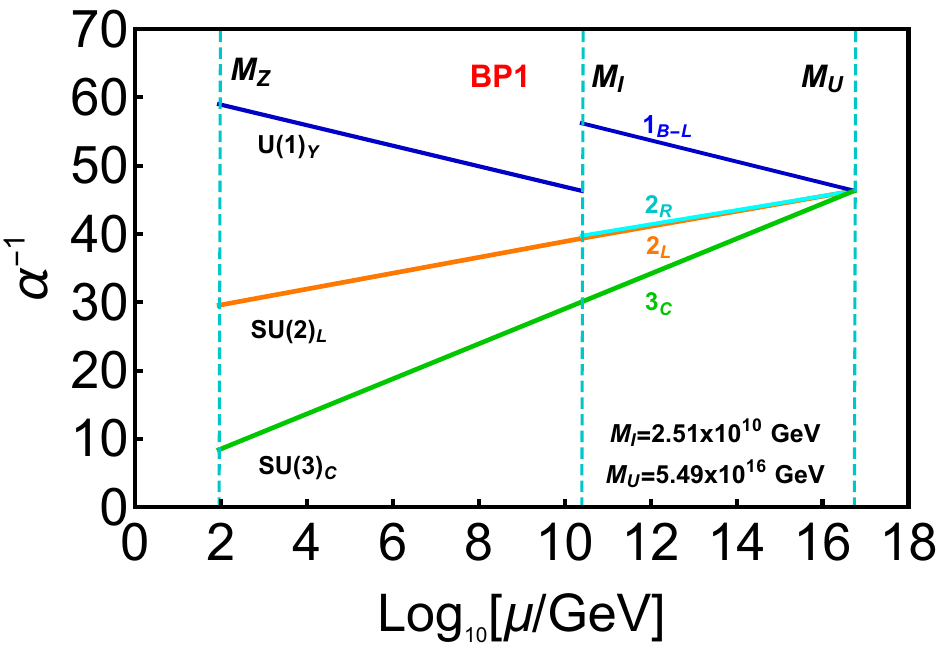}
\includegraphics[scale=0.45]{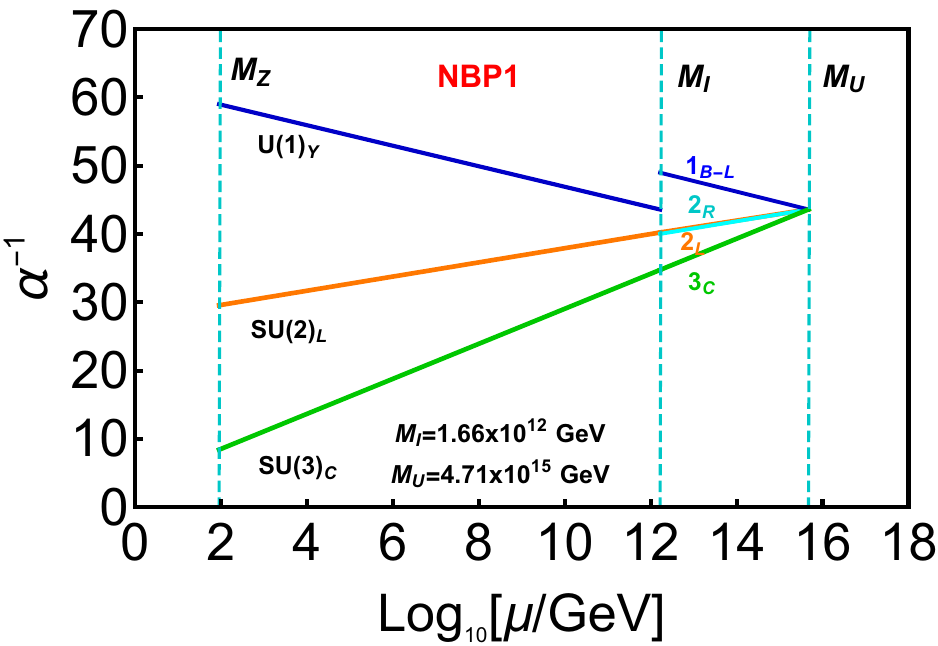}
\includegraphics[scale=0.45]{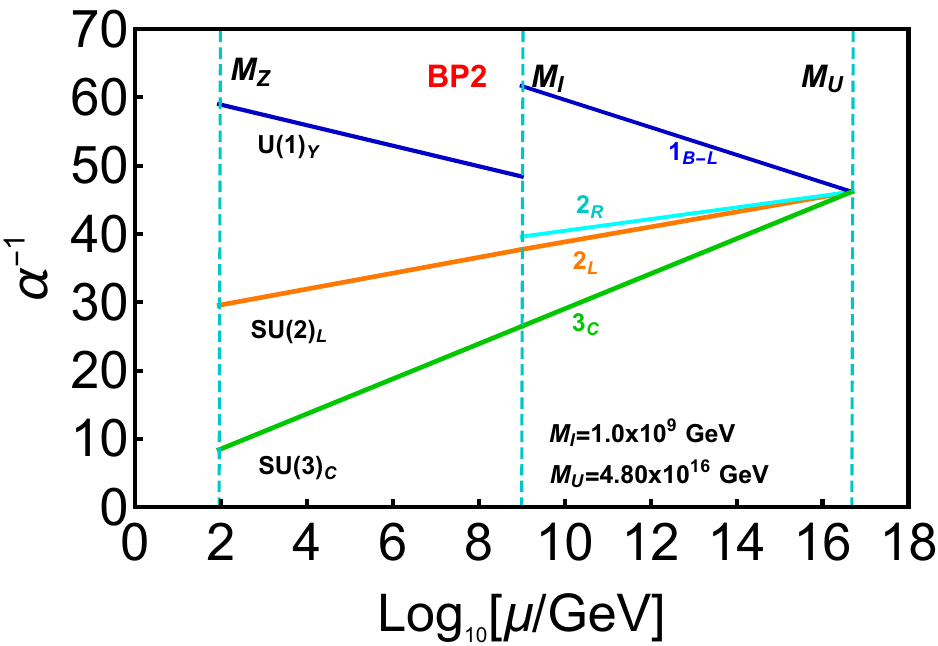}
\includegraphics[scale=0.45]{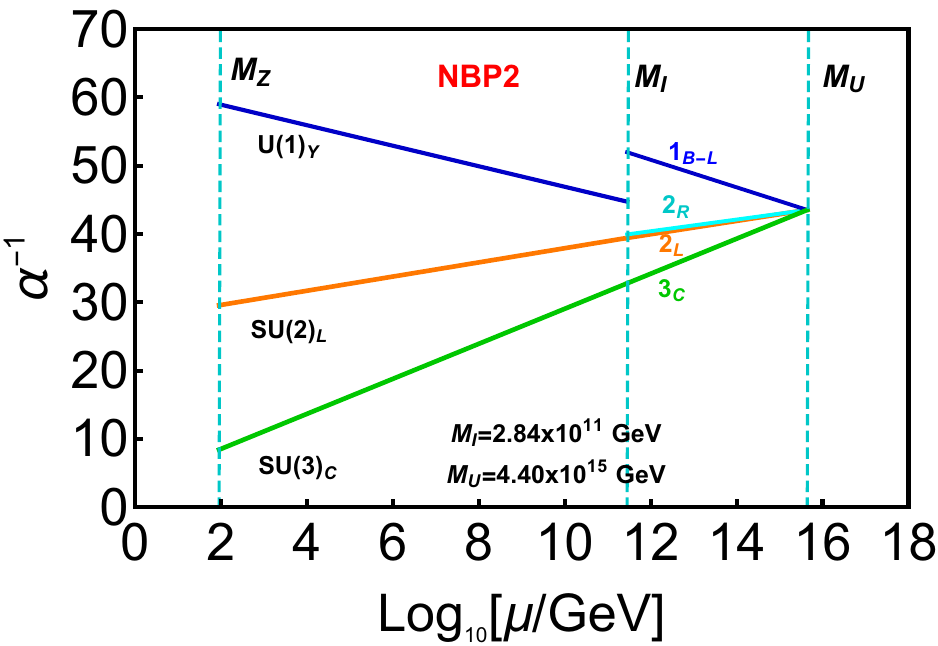}
\caption{Unification plot showing successful gauge coupling unification satisfying proton decay constraints.}
\label{plot:unifn_wo_threshold}
\end{figure*}
\end{widetext}
 
The model predictions on intermediate-mass scale $M_I$, unification mass scale, and inverse GUT scale coupling $\alpha^{-1}_U\equiv \alpha^{-1}_{\rm GUT}$ also depend upon how left-right symmetry is spontaneously broken down to SM. We have considered different benchmark points by considering the breaking of LRSM to SM symmetry via Higgs doublet $H_R$ (BP1), and Higgs triplet $\Delta_R$ (BP2). These choices of scalars needed for spontaneous symmetry breaking of LRSM to SM have been presented below with their respective one-loop beta coefficients. 
\begin{eqnarray}
\mbox{BP1:}\quad &&\mbox{Minimal LRSM: $\Phi (1,2,2,0)$), $H_R(1,1,2,1)$} \nonumber \\
&&\big(b_{3C}, b_{2L}, b_{Y} \big) = \big( -7, -19/6, 41/10 \big) \, 
\nonumber \\ &&
     \big(b'_{3C}, b'_{2L}, b'_{2R}, b'_{BL} \big) = \big( -7, -3, -17/6, 17/4 \big)   \nonumber \\
\mbox{BP2:}\quad &&\mbox{Manifest LRSM: $\Phi (1,2,2,0)$), $\Delta_R(1,1,2,-2)$} \nonumber \\
&&\big(b_{3C}, b_{2L}, b_{Y} \big) = \big( -7, -19/6, 41/10 \big) \, 
\nonumber \\ &&
     \big(b'_{3C}, b'_{2L}, b'_{2R}, b'_{BL} \big) = \big( -7, -3, -7/3, 11/2 \big)   \nonumber 
\end{eqnarray}
 One can consider two bidoublet scalars instead of one bidoublet scalar used in BP1 and BP2 for correct fermion mass fitting in $SO(10)$ GUT. We only estimated the model parameters expressed in terms of one-loop beta coefficients. Using these estimated values of parameters, the values of intermediate left-right symmetry breaking scale $M_I$, unification mass scale $M_U$, and inverse GUT coupling constant $\alpha^{-1}_U$ are estimated and presented in left panel of Fig.(\ref{plot:unifn_wo_threshold}) with successful unification of gauge couplings. From left to right, the vertical dotted lines in Fig.(\ref{plot:unifn_wo_threshold}) show the symmetry-breaking scales $M_Z$ as electroweak scale, $M_I$ as intermediate left-right symmetry breaking scale and $M_U$ as the unification scale. One such estimation correspond to BP2 is given below, 
$$M_I = 10^{9}\,,\quad M_U = 4.80 \times 10^{16}\,,\quad \alpha^{-1}_U = 46.20$$. 

We wish to examine how these model predictions are modified if we include a scalar triplet $\Omega$ with zero hypercharge from electroweak symmetry breaking onwards. We designate these new benchmark points as NBP1 and NBP2 by adding $\Omega$ from the SM symmetry breaking scale onwards and the corrosponding modification is presented in the right panel of Fig.(\ref{plot:unifn_wo_threshold}). Here, we have denoted $B-L$ as $BL$ for notational simplicity.

We consider one benchmark point with unification mass scale $M_U$ and inverse GUT coupling constant as,
$$M_U= 10^{15.65}\,\,\mbox{GeV and} \quad \alpha^{-1}_G=43.5$$
The numerically estimated proton lifetime for this benchmark point is $\tau_p = 4.0 \times 10^{34}$ yrs.
In Table (\ref{tab:pdecay}), we have provided proton decay lifetime $\tau_p$ calculated numerically in the $SO(10)$ GUT with intermediate left-right symmetry. The projected proton lifetimes are provided for BP1, NBP1, and NBP2, all consistent with the Super-Kamiokande or Hyper-Kamiokande experiment's upper limit.

\begin{table}[h]
\centering
\vspace{-2pt}
\begin{tabular}{||c|c|c|c||}
\hline \hline
 Benchmark Points & $M_U$\,\mbox{(GeV)} & $\alpha_G^{-1}$& $\tau_p$ (yrs) \\
\hline
\hline
 BP1 & $5.49 \times 10^{16}$ &$46.34$ 
                               & ${{8.6 \times 10^{38}}}$  \\
\hline
 NBP1 & $4.71 \times 10^{15}$ &$43.60$ 
                               & $\pmb{4.77 \times 10^{34}}$  \\
\hline
 NBP2 & $4.40 \times 10^{15}$ &$43.52$ 
                               &$\pmb{3.5 \times 10^{34} }$  \\
\hline \hline
\end{tabular}
\caption{Numerical estimation of proton decay lifetime $\tau_{p}$ in $SO(10)$ GUT with intermediate left-right symmetry. The estimated proton lifetimes are presented in various benchmark points, which agree with the limit set by the Super-Kamiokande or {Hyper-Kamiokande experiment}.}
\label{tab:pdecay}
\end{table}
 
\section{CDF W boson mass anomaly and Supersymmetric $SO(10)$ GUT}
\label{sec:SO10:Wmass-susy}
\begin{figure}[t]
\centering
\includegraphics[width=0.8\linewidth]{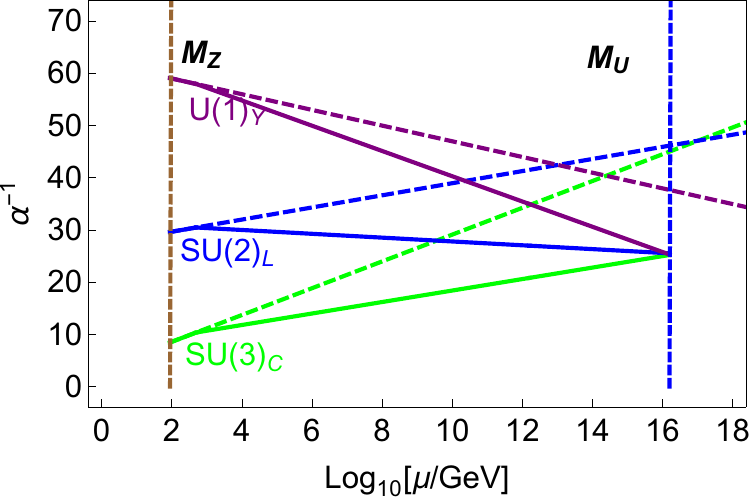}
\caption{Unification plot for MSSM with unification mass scale $M_{U}=10^{16.4}$~GeV, $M_{\rm SUSY}=500$~GeV, $M_Z=91.187$~GeV and $\alpha^{-1}_U \simeq 23$ for SUSY $SO(10)$ GUT without having any scalar triplet and with no intermediate symmetry.}
\label{fig:RG-evolution-susy}
\end{figure}

Supersymmetry (SUSY) models predict that a corresponding fermion exists for every boson and vice versa.
Due to the presence of superpartners of SM particles lying around the electroweak scale, gauge couplings receive corrections from the exchange of superpartners in loop diagrams, leading to a modification of the running of the gauge couplings with energy. For supersymmetric models, 
\begin{eqnarray}
	b_i
	&=&- 3 \mathcal{C}_{2}(G)  + \sum_{R} T(R) \prod_{j \neq i} d_j(R) \,. \nonumber \\
	&=& - 3 \mathcal{C}_{2}(G) + 2 N_G + T(S)
\label{oneloop_a}
\end{eqnarray}
Here, $\mathcal{C}_{2}(G)$ is the quadratic Casimir operator for gauge bosons in their adjoint representation of a given gauge group, and its value for a given gauge group here,
\begin{equation}
	\mathcal{C}_2(G) \equiv 
	\begin{cases} 
		N & \text{if } SU(N), \\
    0 & \text{if }  U(1).
	\end{cases}
\end{equation}
The other parameter $T(R)$ is the trace of the irreducible representation $R$ for a given fermion or scalar with its analytic formula,
\begin{equation}
	T(R) \equiv 
	\begin{cases} 
		1/2 & \text{if } R \text{ is fundamental}, \\
    N   & \text{if } R \text{ is adjoint}, \\
		0   & \text{if } U(1).
	\end{cases}
\end{equation}
Moreover, $d(R)$ is the dimension of a given representation $R$ under all $SU(N)$ gauge groups, excluding the gauge group for which one loop beta coefficient is derived. The number of fermion generation is denoted by $N_G$, and its value is 3. The explicit scalar contribution to one loop beta coefficient is defined by $T(S)$ for a complex scalar $S$. The value of $T(S)$ is $0$ for $b_{3C}$, $1$ for $b_{2L}$ and $3/5$ for $b_{Y}$. Thus, the derived one-loop beta coefficients for MSSM without inclusion of scalar triplet is $b_i =(-3, 1, 33/5)$ with i=3C,2L,Y. Using standard RGEs, the unification mass scale in the minimal supersymmetric standard model is found to be at $2 \times 10^{16}$ GeV as presented in Fig.(\ref{fig:RG-evolution-susy}). Motivated by CDF II W-boson mass anomaly, the inclusion of a real scalar triplet within MSSM spoils the gauge couplings unification when embedded in SUSY $SO(10)$ GUT.

\begin{figure}[t]
\centering
\includegraphics[width=0.98\linewidth]{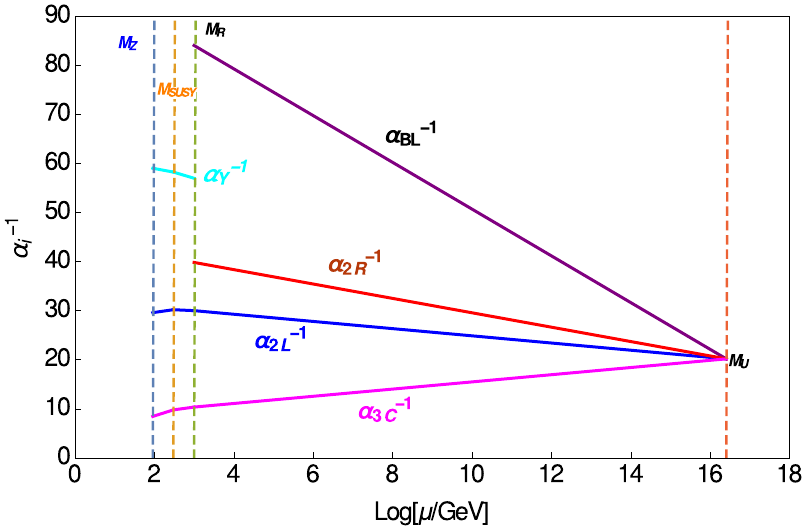}
\caption{Gauge coupling unification plot with $M_{U}=10^{16.4}$~GeV, $M_{R}=5$~TeV, 
$M_{\rm SUSY}=500$~GeV, $M_Z=91.187$~GeV for SUSY $SO(10)$ GUT with intermediate left-right symmetry. Hence, the RH 
Higgs doublets contributions are taken into account in RG evolution of gauge couplings from $M_R$ to $M_U$ while LH doublets are kept at the GUT scale.}
\label{fig:RG-evolution-susyLR}
\end{figure}

We also present a gauge coupling unification plot in Fig.(\ref{fig:RG-evolution-susyLR}) for the SUSY $SO(10)$ GUT having intermediate left-right symmetry motivated for explaining neutrino masses, dark matter, and matter-antimatter asymmetry of the present universe. However, one can introduce intermediate left-right symmetry between MSSM and SUSY $SO(10)$ GUT for addressing successful unification of gauge couplings and proton decay 
constraints~\cite{Ma:1994ge,Hisano:2015pma,Patra:2009wc,Patra:2010ks,Babu:2008ep,Aulakh:1998nn,Kuchimanchi:1993jg,Aulakh:1997fq,Aulakh:1997ba}. With intermediate left-right symmetry, the framework allows $SU(2)_R \times U(1)_{B-L}$ breaking around few TeV scale by non-zero VEV of 
$\overline{16}_H$ resulting TeV scale masses for RH gauge bosons $W_R$, $Z_R$ which can be directly produced at LHC~\cite{BhupalDev:2010he,Malinsky:2005bi,Dev:2009aw}. 
%
%
\section{Summary}
We have studied the non-supersymmetric $SO(10)$ GUT with and without any intermediate symmetry breaking step and its connection to CDF II W boson mass anomaly by including a scalar triplet with zero hypercharge at the electroweak scale. The $SU(2)_L$ scalar triplet is contained in Higgs representation $45_H$ of $SO(10)$ initially introduced to break $SO(10)$ to SM without having any intermediate stage of symmetry breaking. The other representations of $SO(10)$ needed for known usual quark and lepton masses are $10_H$ and $16^{a}_F$ (a=1,2,3) while gauge bosons are lying in the adjoint representation of $45_V$. 

The minimal representations ($45_H$, $16_F$, and $10_H$) of $SO(10)$ GUT, not only explained the reported shift in W boson mass anomaly, but also could account for known SM fermion masses and mixings. Including extra $SU(2)_L$ scalar triplet contained in $45_H$ at electroweak scale onwards helps in achieving unification of gauge couplings corresponding to fundamental forces of SM. However, the predicted unification mass scale, $M_U \simeq 10^{14.4}$~GeV, and the GUT scale inverse fine structure constant, $\alpha^{-1}_U \simeq 39.0$ yield the proton lifetime $\tau_p = 10^{30}\,$ yrs which is well below the current experimental bound by Super-Kamiokande experiments. The issue of proton decay constraints can be avoided if we allow intermediate left-right symmetry between $SO(10)$ to SM. The $SO(10)$ GUT being a rank five group and SM as a rank four group, one can accommodate intermediate symmetry-breaking steps addressing the issue of proton decay constraints, neutrino mass, and other observables which otherwise not possible in case of rank four $SU(5)$ GUT. 

With the intermediate left-right symmetry, the estimated value of the unification mass scale is $M_U \simeq 5 \times 10^{15}$~GeV, proton lifetime as $\tau_p \simeq 5.0 \times 10^{34}$~yrs consistent with experimental bound set by Super-Kamiokande experiment. The additional feature of left-right symmetry at an intermediate scale as to accommodate right-handed neutrinos and/or scalar triplets having non-zero $B-L$ charge which can explain non-zero neutrino mass via Type-I/Type-II seesaw mechanism and address matter-antimatter asymmetry of the universe via decay of right-handed neutrinos or scalar triplets~\cite{Hambye:2005tk,AristizabalSierra:2011ab,AristizabalSierra:2012js,Pramanick:2022put}. 
Alternatively, the impact of two loop contributions, GUT threshold corrections, and gravitational corrections can modify unification mass scale and proton lifetime prediction along with other phenomenology like neutrino mass and matter-antimatter asymmetry of the universe, which will be studied separately. 

In SUSY models, single-stage unification is possible even without including any triplet Higgs scalar. Proton decays are mediated by dimension-4 operators, making the decay rates too fast. This has been solved by imposing R-parity, which has many virtues. However, non-observation of SUSY particles smears out the other features of SUSY models.

\acknowledgments 
Purushottam Sahu would like to acknowledge the Institute Postdoctoral Fellowship of IIT Bombay for financial support. PS also acknowledges the support from the Abdus Salam
International Centre for Theoretical Physics (ICTP) under the 'ICTP Sandwich
Training
Educational Programme (STEP)' SMR.3676 and SMR.3799.

\bibliographystyle{utcaps_mod}
\bibliography{SO10W}
\end{document}